\documentstyle[preprint,aps,eqsecnum]{revtex}
\tighten

\font\twelvemsb=msbm10 at 12pt
\font\ninemsb=msbm7 at 9pt
\font\sixmsb=msbm5 at 6pt
\newfam\Msbfam
\textfont\Msbfam=\twelvemsb
\scriptfont\Msbfam=\ninemsb
\scriptscriptfont\Msbfam=\sixmsb

\def\half{{\textstyle{1\over2}}}
\def\third{{\textstyle{1\over3}}}
\def\twothird{{\textstyle {2\over 3}}}
\def\quar{{\textstyle{1\over4}}}

\def\beq{\begin{equation}}
\def\eeq{\end{equation}}
\def\bi{\begin{itemize}}
\def\ei{\end{itemize}}
\def\beqar{\begin{eqnarray}}
\def\eeqar{\end{eqnarray}}

\newcommand{\rmd}{{\,\rm d\null}}
\newcommand{\tr}{\mathop{\rm tr\null}}

\def\lra{\mathop{\hbox to .5in{\rightarrowfill}}}

\let\varkappa\kappa

\skip\footins = 14pt 
\begin{document}


\title{CHERN-SIMONS REDUCTION AND NON-ABELIAN \\
FLUID MECHANICS\footnotemark[1]}

\author{R.~Jackiw and V.P. Nair\footnotemark[2]}

\address{Center for Theoretical Physics\\ Massachusetts
Institute of Technology\\ Cambridge, MA ~02139--4307,
USA}

\author{So-Young Pi}

\address{Physics Department\\ Boston University\\ Boston,
MA ~02215, USA}

\footnotetext[1] {\baselineskip=12pt This work is supported
in part by funds provided by  the U.S.~Department of Energy
(D.O.E.) under contract
DE-FC02-94ER40818, DE-FG02-91ER40676 and by NSF grant PHY-9605216. 
\break \hfill MIT-CTP-2971, BU HEP-00-06,
hep-th/0004084, April 2000}
\footnotetext[2]{\baselineskip=12pt Permanent address: Physics Department,
City College of the CUNY, New York, NY 10031.}

\maketitle
\begin{abstract}%
We propose a non-Abelian generalization of the Clebsch parameterization
for a vector in three dimensions. The construction is based on a group-theoretical
reduction of the Chern-Simons form on a symmetric space. The formalism is then used to
give a canonical (symplectic) discussion of non-Abelian fluid mechanics, analogous to
the way the Abelian Clebsch parameterization allows a canonical description of
conventional fluid mechanics.

\end{abstract}
\renewcommand{\baselinestretch}{1.15}

\section{Introduction}

In a recent paper \cite{JP1}, a pure $SU(2)$ gauge potential
$A=g^{-1} \rmd g$,
whose Chern-Simons term
\begin{eqnarray}
CS(A)&&= \int \omega (A)\nonumber\\
\omega(A) &&= -\frac{1}{8 \pi^2}  \tr (A \rmd A +\twothird 
A^3) = \frac{1}{16\pi^2} \left( A^a \rmd A^a + \third
\epsilon^{abc} A^a A^b A^c\right)
\label{eq:1.1}
\end{eqnarray}
measures
the quantized winding number of $g$ \cite{DJT},
\beq
CS(g^{-1} \rmd g) = \frac{1}{24\pi^2} \int \tr (g^{-1} \rmd
g)^3 \equiv W(g)
\label{eq:1.2}
\eeq
was used to generate a $U(1)$ potential $a$, by projection
onto an Abelian direction. 
\beq
a=- 2~ \tr t^3 g^{-1} \rmd g
\label{eq:1.3}
\eeq
[$t^a$ are anti-Hermitian generators of the $SU(2)$ group, normalized by
$\tr t^at^b = -\half \delta^{ab}$.]  The Abelian potential $a$ is
not a pure gauge, and it contains three arbitrary functions
[corresponding to the three parameters of $SU(2)$], hence $a$ can
represent an arbitrary Abelian 3-vector.  The Chern-Simons
3-form for $a$ coincides with that of its $SU(2)$
pure gauge antecedent, by virtue of the $SU(2)$ identity $\tr (t^3 g^{-1}
\rmd g) \rmd \tr (t^3 g^{-1} \rmd g) = (1/3!)\tr (g^{-1} \rmd
g)^3$.  Thus the constructed Abelian potential possesses
quantized Chern-Simons number (or magnetic helicity)
\cite{wolt} $(1/16\pi^2)\int
a \rmd a$, equal to $W(g)$, the winding number of $g$. 
Because within $SU(2)$, $\tr (g^{-1} \rmd g)^3$ can be
explicitly presented as a total divergence \cite{4}, the Abelian
Chern-Simons density $a\rmd a$ also appears as a total
divergence.  This in turn indicates that the Clebsch
parameterization for $a$
\beq
a=\rmd \theta + \alpha \rmd \beta
\label{eq:1.4}
\eeq
can be readily constructed.
This parameterization of an Abelian potential $a$ ensures
that the corresponding Chern-Simons density $a\rmd a$ is a total
divergence.
\begin{eqnarray}
a\rmd a = \rmd \theta \rmd \alpha \rmd \beta &=&
\rmd (\theta \rmd \alpha \rmd \beta) \nonumber  \\
&=&
-\rmd (\rmd \theta \alpha \rmd \beta) \nonumber \\
&=&
\rmd (\rmd \theta \rmd \alpha \beta)  
\label{eq:1.5}
\end{eqnarray}

In this paper we discuss how the above structures extend to
the non-Abelian situation. Also we use our non-Abelian quantities to construct a canonical
theory of non-Abelian fluid mechanics, analogous to the way in which the Abelian Clebsch
parameterization is used in ordinary fluid mechanics.

We begin, in section II, with a pure gauge $g^{-1}\rmd g$ in
some non-Abelian group $G$ (called the Ur-group),
and the Chern-Simons term again coincides
with the winding number of $g$ as in
(\ref{eq:1.1}), (\ref{eq:1.2}).  We consider a normal subgroup $H$, with
generators $I^\alpha$, and construct a non-Abelian gauge field by
projection.
\beq
A^\alpha \propto ~\tr (I^\alpha g^{-1} \rmd g )
\label{eq:1.6}
\eeq
Within $H$, this is not a pure gauge.
We determine the group structure that is needed to ensure
that the Chern-Simons 3-form $\omega (A)$ of $A^\alpha$
is proportional to 
$\tr (g^{-1} \rmd g)^3$, so that the Chern-Simons number of
$A^\alpha$ equals the winding number of $g$.  In this way we
construct non-Abelian gauge fields, belonging to the group
$H$, with quantized Chern-Simons number.  Moreover, we
describe the properties of the Ur-group $G$,
that are needed so that the
projected potential $A^\alpha$ enjoys sufficient generality to
represent an arbitrary potential in $H$.

Since $\tr (g^{-1} \rmd g)^3$ is a total derivative for an
arbitrary group (although this fact cannot in general be
expressed in finite terms) \cite{cron} our construction ensures that the
form of $A^\alpha$, which is achieved through the projection (\ref{eq:1.6}),
produces a total derivative expression for its Chern-Simons density
$\omega (A)$. 

With the above mentioned properties for the potential,
it is appropriate to consider (\ref{eq:1.6}) as a
``non-Abelian Clebsch parameterization".

In explicit examples, which we present in section III, it is found that the
``total derivative" form for the Chern-Simons density of
$A^\alpha$ is
achieved in two steps.  The parameterization (\ref{eq:1.6})
directly leads to an Abelian form of the Chern-Simons
density
\beq
A^\alpha \rmd A^\alpha + \third f^{\alpha \beta \gamma}
A^\alpha A^\beta A^\gamma  =
\gamma \rmd \gamma
\label{eq:1.7}
\eeq
for some $\gamma$.  Then Darboux's theorem \cite{J2} (or usual fluid
dynamical theory \cite{lamb}) ensures that $\gamma$ can be presented
in Clebsch form, so that $\gamma \rmd \gamma$, is
explicitly a total derivative.

We also observe that at least for
$SU(2)$ one can do without the above general discussion
and directly present a parameterization for arbitrary
$SU(2)$ potentials, which produces a total derivative
expression for the $SU(2)$ Chern-Simons density.  The
parameterization is a natural generalization into the
non-Abelian context of the Clebsch parameterization (\ref{eq:1.4}), which
achieves the total derivative form for Abelian Chern-Simons
densities.

In section IV, we construct a non-Abelian version of
fluid mechanics and
magnetohydrodynamics, which may be useful
as an effective description for the long wavelength
degrees of freedom
in a quark-gluon plasma.
\section{Parameterization of potentials and\\ the
Chern-Simons 3-form}
\subsection{General Considerations}

As stated in the introduction, we consider the
parameterization of gauge potentials for a group $H$ of the
form $\tr (I^\alpha g^{-1} \rmd g)$, where $g$ is an
element of a group $G$, $H$ being a subgroup of $G$ and
$I^\alpha$ are the generators of $H$.  Conditions on the Ur-group
$G$, which we take to be compact and semi-simple, are
the following. First of all $G$ has to
be so chosen that it has sufficient number of parameters to
make  $\tr (I^\alpha g^{-1} \rmd g)$ a generic potential for
$H$.  Since we are in three dimensions, an $H$-potential
$A^\alpha_i$ has $3\times {\rm dim} H$ independent functions; so a
minimal requirement will be
\beq
{\rm dim~} G \ge 3 {\rm ~dim~} H \ \ .
\label{eq:2.1}
\eeq
Secondly we require that the $H$-Chern-Simons form for $A^\alpha$
should coincide with that of $g^{-1}\rmd g$, thereby ensuring that
the $H$-potential possesses (quantized) Chern-Simons number
equal to $W(g)$
and also that the $H$-Chern-Simons density is a total derivative, or at least can be
brought to an Abelian form as in (\ref{eq:1.7}).
As we shall show in a moment, this is achieved if $G/H$ is a
symmetric space.  In this case, if we split the Lie algebra of
$G$ into the $H$-subalgebra spanned by $I^\alpha$,
$\alpha=1,\dots, {\rm dim~} H$, and the orthogonal
complement spanned by $S^A$, $A=1,\dots, ({\rm
dim~}G-{\rm dim~}H)$, the commutation rules are of the
form
\begin{mathletters}%
\begin{eqnarray}
\relax [I^\alpha, I^\beta{]} &=& f^{\alpha \beta \gamma}
I^\gamma
\label{eq:2.2a} \\
\relax[ I^\alpha, S^A{]} &=& h^{\alpha AB} S^B
\label{eq:2.2b} \\
\relax[S^A, S^B{]} &=& c ~h^{\alpha AB} I^\alpha
\label{eq:2.2c}
\end{eqnarray}
\end{mathletters}%
$(h^\alpha)^{AB}$ form a (possibly reducible)
representation of the
$H$-generators $I^\alpha$. The constant  $c$
depends on 
normalizations. More explicitly, if the structure
constants for the Ur-group $G$ are named ${\bar f}^{abc},~
a,b,c =1,\dots,{\rm dim}G$, then the conditions (\ref{eq:2.2a}-c) require
that ${\bar f}^{abc}$ vanishes whenever an odd number of indices belongs to
the orthogonal complement labeled by $A,B,..$. Moreover, $f^{\alpha \beta \gamma}$ 
are taken to be the conventional structure constants for $H$ and this may
render them proportional to (rather than equal to) ${\bar f}^{\alpha \beta \gamma}$.

We define the
traces of the generators by
\begin{eqnarray}
\tr (I^\alpha I^\beta) &=& -a~ \delta^{\alpha\beta} \ \ ,
\quad \tr (S^A S^B) = -b~ \delta^{AB}
\nonumber  \\
\tr (I^\alpha S^A) &=&0 \ \ .
\label{eq:2.3}
\end{eqnarray}
We can evaluate the quantity $\tr[S^A,S^B]I^\alpha =
\tr S^A[S^B,I^\alpha]$ using the commutation rules and
(\ref{eq:2.3}).  This immediately gives the relation
$ac=b$. 

Expanding $g^{-1} \rmd g$ in terms of generators, we write
\begin{equation}
g^{-1}  \rmd g= (I^\alpha A^\alpha + S^A\alpha^A)
\label{eq:2.4}
\end{equation}
which defines the $H$-potential $A^\alpha$. 
Equivalently
\begin{equation}
A^\alpha =-\frac{1}{a} \tr(I^\alpha g^{-1}  \rmd g)
\label{eq:2.5}
\end{equation}
(Such projected potentials have been used before in formulating sigma models,
see \cite{bal}).
From $\rmd (g^{-1} \rmd g)=-g^{-1} \rmd g~  g^{-1} \rmd g$, we get the
Maurer-Cartan relations
\begin{eqnarray}
F^\alpha\equiv \rmd A^\alpha + \half
f^{\alpha\beta\gamma}A^\beta A^\gamma &&= -\frac{c}{2}
h^{\alpha A B} \alpha^A \alpha^B  \nonumber \\
\rmd \alpha^A + h^{\alpha B A} A^\alpha \alpha^B &&= 0 \ \ .
\label{eq:2.6}
\end{eqnarray}
Using these results, the following chain of equations shows that
the
Chern-Simons 3-form for the $H$-gauge group is
proportional to $\tr (g^{-1} \rmd g)^3$.
\begin{eqnarray}
\omega (A)&=& \frac {1}{16\pi^2} (A^\alpha
\rmd A^\alpha+\third f^{\alpha\beta\gamma}A^\alpha
A^\beta A^\gamma) \nonumber \\
&=&\frac{1}{48\pi^2} ( A^\alpha
\rmd A^\alpha+2~ A^\alpha F^\alpha) \nonumber \\
&=&\frac {1}{48\pi^2} ( A^\alpha \rmd A^\alpha+c~ 
h^{\alpha AB}A^\alpha \alpha^A\alpha^B) \nonumber \\
&=&\frac {1}{48\pi^2} ( A^\alpha
\rmd A^\alpha+c~ \rmd \alpha^A \alpha^A)   \nonumber\\
&=&-\frac
{1}{48\pi^2}[{1\over a} \tr(A \rmd A)+\frac{c}{b}\tr(\rmd
\alpha\alpha)]
 \nonumber\\
&=&-\frac {1}{48\pi^2 a} \tr(A\rmd A + \alpha \rmd  \alpha)
\nonumber\\ 
&=&-\frac {1}{48\pi^2a} \tr g^{-1} \rmd g  ~\rmd
(g^{-1}
\rmd g) \nonumber \\ 
&=&\frac {1}{48\pi^2a} \tr(g^{-1} \rmd g)^3
\end{eqnarray}
\label{eq:2.7}
In the
above sequence of manipulations, 
we have used the Maurer-Cartan relations
(\ref{eq:2.6}), which rely on the symmetric space structure of
(\ref{eq:2.2a}-c), and the trace relations (\ref{eq:2.3}), along with $ac=b$.

We thus see that $\int \omega (A)$ is indeed the winding number of
the configuration $g\in G$.  Since $\tr(g^{-1} \rmd g)^3$ is a
total derivative locally on $G$, the potential (\ref{eq:2.5}),
with the symmetric space structure of (\ref{eq:2.2a}-c), does indeed obey
the requirement of making $\omega (A)$ a total derivative. 
It is therefore appropriate to call our construction (\ref{eq:2.5})
a ``non-Abelian Clebsch parameterization".

\subsection{Choosing the Ur-group \protect\boldmath$G$}

In explicit realizations, given a gauge group of interest $H$,
we need to choose a group $G$ 
such that the conditions
(\ref{eq:2.1}), (\ref{eq:2.2a}-c) hold. In general this is not possible.
However, one can proceed recursively. Let us suppose that the
desired result has been established for a group, which we call
$H_2$. Then we form $H\subset G$ obeying (\ref{eq:2.2a}-c)
as $H=H_1 \times H_2$, where
$H_1$ is the gauge group of interest, satisfying ${\rm dim}G \geq
3~{\rm dim} H_1$. For this choice of $H$, the result (\ref{eq:2.7})
becomes
\begin{eqnarray}
\omega (H_1) +\omega (H_2) =  \frac {1}{48\pi^2a}\tr(g^{-1}
\rmd g)^3
\label{eq:2.8}
\end{eqnarray}
But since $\omega (H_2)$ is already known to be a total derivative,
(\ref{eq:2.8}) shows the desired result: $\omega (H_1)$ is a total
derivative.

As a specific example, consider the orthogonal groups $O(n)$
for which we can use $G= O(2n-1)$ and $H= O(n) \times O(n-1)$.
The case $n=2$, with $O(1)={\bf 1}$, reproduces the previous results
of the Abelian construction $O(3)$ [or $SU(2)$]$ \rightarrow O(2)$ \cite{JP1}.
For $n=3$, $G$ is $O(5)$, $H_1= O(3)$ and $H_2= O(2)$.
Since $\omega [O(2)]$ is already known to be a total derivative, we
learn from (\ref{eq:2.8}) that the Chern-Simons density
for $O(3) ~[SU(2)]$ is also a total derivative.
(Explicit formulas for this case are presented in the next section.)
Evidently the procedure can be continued for arbitrary $O(n)$, but we have
not found a simple sequence of embeddings for other groups.

To see that the algebra of $G= O(2n-1)$ and $H= O(n)\times O(n-1)$
satisfies (\ref{eq:2.2a}-c) we proceed as follows.
Let $\Gamma^\alpha$ denote the set of
Dirac gamma matrices in $n$ dimensions and let $\gamma^i$ denote
the set of Dirac gamma matrices in $(n-1)$ dimensions.
These are considered as acting on different vector spaces.
We have
\begin{eqnarray}
\Gamma^\alpha\Gamma^\beta + \Gamma^\beta
\Gamma^\alpha = 2~ \delta^{\alpha\beta} \nonumber \\
\gamma^i \gamma^j + \gamma^j \gamma^i  = 2~\delta^{ij} 
\label{eq:2.9}
\end{eqnarray}
We construct
\begin{eqnarray}
\Sigma^{\alpha\beta} &=& \frac{i}{2} [\Gamma^\alpha , \Gamma^\beta] \nonumber
\\
\sigma^{ij} &=& \frac {i}{2} [\gamma^i , \gamma^j ] \label{eq:2.10} \\
S^{\alpha i} &=& \Gamma^\alpha \gamma^i 
\nonumber
\end{eqnarray}
$\Sigma^{\alpha\beta}$ are (Hermitian) generators of $O(n)$, $\sigma^{ij}$ are
generators of $O(n-1)$.  The set of matrices
$(\Sigma^{\alpha\beta}, \sigma^{ij}, S^{\alpha i})$ form the generators of
$O(2n-1)$.  From (\ref{eq:2.10}) we find that
$S^{\alpha i}$ is a vector of $O(n)$ and a vector of $O(n-1)$ and also
that
\begin{equation}
i~[S^{\alpha i}, S^{\beta j}] = 2~ \delta^{ij}  ~\Sigma^{\alpha\beta} + 2~
\delta^{\alpha\beta}~ \sigma^{ij}
\label{eq:2.11}
\end{equation}
We have thus the required structure (\ref{eq:2.2a}-c).  We construct the
$O(n)$-gauge potential as
\begin{equation}
A^{\alpha\beta}=\frac{i}{a} \tr(\Sigma^{\alpha\beta} g^{-1}  \rmd g)
\label{eq:2.12}
\end{equation}
The number of arbitrary functions present in $A^{\alpha \beta}$ defined by
(\ref{eq:2.12}) is ${3\over 2}n(n-1)$, which is exactly the right number
for an $O(n)$-gauge potential in three dimensions, so that (\ref{eq:2.1})
is satisfied in just the right way (as an equality).
This result is seen as follows. The $O(2n-1)$ group element
$g$ depends on $(n-1)(2n-1)$ parameters. However the trace with 
$\Sigma^{\alpha\beta}$ removes dependence on the $\half (n-1)(n-2)$
parameters of the $O(n-1)$ subgroup. This is a consequence of the fact that
$A^{\alpha \beta}$ is unchanged when $g$ is replaced by $gk$, with 
$k\in O(n-1)$.
\begin{eqnarray}
\tr (\Sigma^{\alpha\beta}g^{-1}\rmd g ) &&\rightarrow \tr [ \Sigma^{\alpha\beta}
(gk)^{-1} \rmd (gk) ]\nonumber\\
&&= \tr [\Sigma^{\alpha\beta} (k^{-1} (g^{-1} \rmd g) ~k +k^{-1}\rmd k )]\nonumber\\
&&= \tr (k \Sigma^{\alpha \beta} k^{-1} g^{-1} \rmd g ) +\tr (\Sigma^{\alpha\beta}
k^{-1}\rmd k)\label{eq:2.12a}
\end{eqnarray} 
The second term on the right vanishes due to the orthogonality of traces
of $O(n)$ with $O(n-1)$ generators, while $k$ disappears from
the first since it commutes with $\Sigma^{\alpha\beta}$.

In the above construction, one can also define an $O(n-1)$-potential
\begin{equation}
A^{ij}=\frac{i}{a} \tr(\sigma^{ij} g^{-1} \rmd g)
\label{eq:2.13}
\end{equation}
Indeed, this is the potential that enters 
$\omega (H_2)=\omega [O(n-1)]$.  This potential
depends on the functions used to construct
$A^{\alpha\beta}$ in (\ref{eq:2.12}).  Thus (\ref{eq:2.13}) does not give
an independent $O(n-1)$-potential. But this is immaterial
since we are really interested in the $O(n)$-potential; (\ref{eq:2.13})
enters our discussion only in the formula for the Chern-Simons density,
namely
$\omega [O(n)] =(1/48\pi^2 a){\tr(g^{-1} \rmd g)^3} -
\omega [O(n-1)]$.

As we have already noted, the potential (\ref{eq:2.12}) depends on just the
right number of arbitrary functions. We shall now show explicitly
that it is sufficiently general to reproduce an arbitrary $O(n)$-gauge
potential that lies close to the trivial gauge orbit 
$A^\alpha =0$ (or $A^\alpha=$ pure gauge).
[For compactness we rename $A^{\alpha\beta}$ of (\ref{eq:2.12})
simply as $A^\alpha$.]
Potentials in the neighborhood of the trivial
gauge orbit may be obtained by writing
$g=\exp(iS_A \theta^A)\cdot h\,k$, $h \in H_1=O(n)$,
$k\in H_2=O(n-1)$. Expanding
in powers of $\theta^A$, we then find
\begin{eqnarray}
A^\alpha &=& R^{\alpha\gamma} (h) a^\gamma + (h^{-1}\rmd h)^\alpha
\nonumber \\
a^\gamma &\approx& {c\over 2} h^{\gamma AB} \theta^B \rmd  \theta^A + \cdots
\label{eq:2.14}
\end{eqnarray}
where $R^{\alpha\gamma} (h)$ is defined by 
\begin{equation}
h I^\alpha h^{-1}=R^{\alpha\gamma}(h)I^\gamma
\label{eq:2.14a}
\end{equation}
($k\in H_2$ drops out of the expression for the potential as discussed earlier.)
Eq.~(\ref{eq:2.14}) tells us that
$A^\alpha$ is the gauge transform of the potential $a^\gamma$.
For small $\theta$'s, this can be brought to the Clebsch form for each
value  of the Lie algebra index $\gamma$. 
We can see this as follows.
There are $n(n-1)=2 {\rm dim} H_1$ functions $\theta^A$ in the expression for
$a^\gamma$. [Additional $\half n (n-1)$ parameters are contained in $h$, giving the total
of ${3\over 2}n(n-1)$ parameters for the potential (\ref{eq:2.12}).] 
$h^{\gamma AB}$
is antisymmetric in $A,B$.
By choosing an appropriate basis one can present the commutator (\ref{eq:2.2c}) in the form
$[S^{(\gamma )},{\tilde S}^{(\gamma )}]\propto I^\gamma$, $\gamma =1,2,\dots,{\rm
dim}H_1=\half n (n-1)$. (There is no summation over $\gamma$. $S^{(\gamma )},~{\tilde
S}^{(\gamma )}$ are selected linear combinations of the $S^A$'s.) In this basis, for each
$\gamma$,
$a^\gamma \approx \alpha^{(\gamma )}\rmd \beta^{(\gamma )}$, (no summation over
$\gamma$), where $\alpha^{(\gamma )}$ and $\beta^{(\gamma )}$ are independent
combinations of the $\theta^A$'s.
This manifestly display s$a^\gamma$ in the Clebsch form for
each value of $\gamma$.
Since we know that any vector in three 
dimensions can be brought to the Clebsch form, Eq.~(\ref{eq:2.14}) tells
us that any gauge potential,
which is sufficiently close to the trivial one, 
can be brought to the form (\ref{eq:2.5}).
In other words, (\ref{eq:2.5}) [or (\ref{eq:2.12})] is a general
parameterization for gauge potentials in a small neighborhood of
$A=0$ (or pure gauge)
in the space ${\cal A}$ of three-dimensional gauge potentials. Since
${\cal A}$ is an affine space, it may be possible to extend this result 
over a larger neighborhood.
A different way of stating this result is as follows.
The arbitrary functions appearing in the expression for the potential, namely, the
gauge parameters contained in $h$ and the coset parameters $\theta^A$, give a choice of coordinates
on ${\cal A}$. This choice of coordinates is valid near the trivial gauge orbit
or near the origin in the gauge-invariant configuration space ${\cal C}={\cal A}/{\cal G}$,
the space of gauge potentials modulo gauge transformations.

It is a well-known theorem, in the context of universal connections, 
that any gauge potential can be written in the form
(\ref{eq:2.5}) for a sufficiently large group $G$ \cite{nara}. 
In general, this requires ${\rm dim}G \geq (d+1)(2d+1)({\rm dim}H)^3$
for gauge potentials of unitary groups in d dimensions.
(The case of orthogonal groups can be realized 
as a special case of the unitary one, and a similar condition
on the dimensions holds.)
It is interesting to note that we have a parameterization of the gauge potential
with the minimal number of parameters, namely, $3~{\rm dim}H$, which
is significantly smaller than what appears in the 
construction of universal connections. It may be that our parameterization does not capture
all the topological subtleties that gauge fields in three dimensions
can have. It should also be pointed out that any parameterization,
and not just ours, has drawbacks. This is because the configuration space
${\cal C}$, for non-Abelian groups, has nontrivial topology and hence
one cannot choose coordinates globally valid on
${\cal C}$. (In the Abelian case, ${\cal C}$ is topologically trivial
for fields on ${\bf R}^3$ and globally valid parameterizations exist.)

\section{The \protect\boldmath$O(3)$ Gauge Potential}

We take $G=O(5), H=O(3)\times O(2)$.  We consider the
4-dimensional spinorial representation of $O(5)$.
With the generators normalized  by $\tr(t^at^b) = -
\delta^{ab}$, the Lie algebra generators of $O(5)$ are given by
\begin{eqnarray}
I^\alpha &=&{1\over 2i} 
\left(
\begin{array}{cc}
\sigma^\alpha & 0 \\
0 &\sigma^\alpha
\end{array}
\right)  \nonumber \\
I^0 &=&{1\over 2i} 
\left(
\begin{array}{cc}
-1 & 0 \\
0 &1
\end{array}
\right)  \label{eq:3.1}   \\
S^A &=&{1\over i\sqrt{2}}
\left(
\begin{array}{cc}
0 & 0 \\
\sigma^A &0
\end{array}
\right)
\qquad
\tilde{S}^A = {1\over i\sqrt{2}}
\left(
\begin{array}{cc}
0 & \sigma^A \\
0 &0
\end{array}
\right)
\nonumber
\end{eqnarray}
$\sigma$'s are the $2 \times 2$ Pauli matrices. $I^\alpha$ generate
$O(3)$, with the conventional structure constants $\epsilon^{\alpha\beta\gamma}$,
and $I^0$ is the generator of $O(2)$.  $S,\tilde{S}$ are the coset
generators. 

A general group element in $O(5)$ can be written in the form $g=M~ hk$
where $h\in O(3)$, $k\in O(2)$, and
\begin{equation}
M = \frac{1}{\sqrt{1+{\bf \bar{w}} \cdot {\bf w} - {1\over 4}({\bf w}\times{\bf {\bar w}})^2 }}
\left(
\begin{array}{cc}
1- {i\over 2}({\bf w} \times {\bf \bar{w}})\cdot {\bf \sigma} & -{\bf w} \cdot {\bf \sigma}
\\[2ex]
{\bf \bar{w}} \cdot {\bf \sigma} & 1+ {i\over 2}({\bf w} \times {\bf\bar{w}})\cdot {\bf \sigma}
\end{array}
\right)
\label{eq:3.2}
\end{equation}
$w^\alpha$ is a complex 3-dimensional vector, with the bar denoting complex
conjugation. 
${\bf w}\cdot {\bf
\bar{w}} = w^\alpha \bar{w}^\alpha$ and $({\bf w} \times {\bf\bar{w}})^\alpha =
\epsilon^{\alpha\beta\gamma} w^\beta \bar{w}^\gamma$.  The gauge
potential given by $- \tr (I^\alpha g^{-1} \rmd g)$ reads
\begin{eqnarray}
A^\alpha &=& R^{\alpha \beta} (h) ~a^\beta  + (h^{-1} \rmd h)^\alpha
\nonumber \\
a^\alpha &=& \frac{1}{1+ {\bf w} \cdot {\bf\bar{w}} - {1\over 4}({\bf w} \times
{\bf\bar{w}})^2}
\Bigg\{ \frac{w^\alpha \rmd {\bf\bar{w}} \cdot ({\bf w} \times
{\bf\bar{w}}) + \bar{w}^\alpha \rmd {\bf w} \cdot ({\bf \bar{w}}\times {\bf w})}{2} \label{eq:3.3}\\
&&~~~~~~~~~~~~~~~~~~~~~~~~~~~~~~~~~~~~~~~~~~~~~~~~{}+
\epsilon^{\alpha\beta\gamma} (\rmd w^\beta \bar{w}^\gamma -
w^\beta \rmd \bar{w}^\gamma) \Bigg\}
\nonumber
\end{eqnarray}
$A^\alpha$ is the $h$-gauge transform of $a^\alpha $ which depends
on six parameters $(w^\alpha, \bar{w}^\alpha)$.  The three gauge
parameters of $h\in O(3)$,
along with the six  $(w^\alpha, \bar{w}^\alpha)$, give the
nine functions needed to parameterize a general $O(3)$- [or $SU(2)$-] 
potential
in three dimensions.  The Chern-Simons form is 
\begin{eqnarray}
\omega (A) &=& \frac{1}{16\pi^2} (A^\alpha \rmd A^\alpha + \third
\epsilon^{\alpha\beta\gamma} A^\alpha A^\beta A^\gamma)
\nonumber \\
&=&  \frac{1}{16\pi^2} (a^\alpha \rmd a^\alpha + \third
\epsilon^{\alpha\beta\gamma} a^\alpha a^\beta a^\gamma) -
\rmd \left[\frac{1}{16\pi^2}  (\rmd h h^{-1})^\alpha a^\alpha )\right] \nonumber\\
&&\qquad{}+
\frac{1}{24\pi^2} \tr (h^{-1} \rmd h)^3
\label{eq:3.4}
\end{eqnarray}
The second equality reflects the usual response of the Chern-Simons density
to gauge transformations.
Using the explicit form of
$a^\alpha$ as given in (\ref{eq:3.3}), we can further reduce this.
Indeed we find that
\begin{equation}
a^\alpha \rmd a^\alpha + \third
\epsilon^{\alpha\beta\gamma} a^\alpha a^\beta a^\gamma =
(-2) \frac{({\bf \bar{w}} \times \rmd {\bf\bar{w}}) \cdot {\bf\rho} + ({\bf w} \times \rmd
{\bf w}) \cdot {\bf\bar{\rho}}}{[1+ {\bf w}\cdot {\bf \bar{w}} -{1\over 4} ({\bf w} \times {\bf{\bar w}})^2 ]^2}
\label{eq:3.5}
\end{equation}
$$
\rho_k = \half \epsilon_{ijk} \rmd \bar{w}^i \rmd \bar w^j
$$
Defining an Abelian potential
\begin{equation}
a = \frac{{\bf w} \cdot {\rmd}{\bf\bar{w}} -{\bf \bar{w}} \cdot \rmd {\bf w}}{1+ {\bf w}\cdot {\bf \bar{w}}
- {1\over 4}({\bf w} \times {\bf\bar{w}})^2}
\label{eq:3.6}
\end{equation}
we can easily check that $a \rmd a$ reproduces (\ref{eq:3.5}).  In other
words
\begin{equation}
\omega (A) = \frac{1}{16\pi^2}  a\rmd a + \rmd \left[ \frac{ (\rmd
h h^{-1})^\alpha a^\alpha )}{16 \pi^2} \right] + \frac{1}{48\pi^2} \tr(h^{-1} \rmd h)^3
\label{eq:3.7}
\end{equation}
If desired, the Abelian potential $a$ can now be written in the Clebsch form making
$a \rmd a$ into a total derivative.

The
$O(3)$-potential (\ref{eq:3.3}) can also be written in a more compact form
as
\begin{equation}
a^\alpha =\frac{2i}{(1 + 2 \bar{\xi} \cdot \xi +N^2)} \Bigl\{ \rmd
\bar{\xi} J^\alpha \xi - \bar{\xi} J^\alpha \rmd \xi - NJ^\alpha \rmd N
\Bigr\}
\label{eq:3.12}
\end{equation}
where $N^\alpha = \bar{\xi} J^\alpha \xi$, $w_\beta = \sqrt{2}~
\xi_\beta$.  $(J^\alpha)_{\beta\gamma} = -i
\epsilon^{\alpha\beta\gamma}$ is the adjoint representation of the
Lie algebra of $SU(2)$.  

The Abelian gauge potential obtained in \cite{JP1}
by projection from $SU(2)$, in other words the potential (\ref{eq:1.3}),
can also be written in a form very similar to the above expression for
$a^\alpha$.
With $g$ parameterized as
\beq
g=\left( \matrix{\sqrt{1+({\bar \xi}\xi )^2}&\sqrt{2}~\xi\cr
                -\sqrt{2}~{\bar \xi} &\sqrt{1+({\bar \xi}\xi )^2}\cr}\right)\cdot e^{-i\sigma^3\theta /2}
\label{eq:3.12a}
\eeq
the projection (\ref{eq:1.3}) gives
\beq
a =\rmd \theta + \frac{2i}{[1 + 2 \bar{\xi} \xi +(\bar{\xi} \xi)^2]} ( \rmd
\bar{\xi} \xi - \bar{\xi} \rmd \xi )
\label{eq:3.13}
\eeq
where $\xi$ is now just a single complex function. [The Clebsch parameters
$\alpha ,~\beta$ are given by $\xi = \sqrt{\rho} e^{i\beta}$, $\alpha = 4\rho /(1+\rho )^2$.~]

At least for the case of $O(3)$ [or $SU(2)$], there is another way of
parameterizing the potentials, without considering embeddings in a
larger group. This also leads to the reduction of the
Chern-Simons form as
above.  The key observation is the following.  We can write
\begin{equation}
\rmd \chi^\alpha = -\half \epsilon^{\alpha\beta\gamma} \chi^\beta \chi^\gamma 
\label{eq:3.8}
\end{equation}
for $\chi^\alpha =i \tr (\sigma^\alpha g^{-1} \rmd g)$, $ g \in SU(2)$.
Further, $SU(2)$
being three-dimensional, we have
\begin{eqnarray}
\chi^1 \chi^2 \chi^3 &=&\frac{1}{3!}
\epsilon^{\alpha\beta\gamma}
\chi^\alpha \chi^\beta \chi^\gamma \nonumber  \\
&=& -\chi^1 \rmd \chi^1 =-\chi^2\rmd \chi^2 =-\chi^3 \rmd \chi^3 
\label{eq:3.9}
\end{eqnarray}
We take each Lie algebra component of the potential to be proportional
to $\chi^\alpha$, 
\begin{equation}
A^\alpha = f_\alpha \chi^\alpha
\label{eq:3.10}
\end{equation}
with no summation over $\alpha$, i.e., Eq.(\ref{eq:3.10}) holds for
each component separately.  Using (\ref{eq:3.9}), (\ref{eq:3.10}) we then
find
\begin{eqnarray}
\omega (A) &=&\frac{1}{16\pi^2}
(2 f_1f_2f_3 - f^2_1-f^2_2-f^2_3) \chi^1 \chi^2
\chi^3 \nonumber
\\
&=& {1\over 16\pi^2} (f^2-2f_1 f_2 f_3) \chi^3 \rmd \chi^3\nonumber\\
&=&\frac{1}{16\pi^2} a \rmd a
\label{eq:3.11}
\end{eqnarray}
where $a=\sqrt{f^2 -2f_1f_2f_3} ~\chi^3$.  We thus get the Abelian
form again for a suitably defined Abelian potential $a$.  $A^\alpha$ of
(\ref{eq:3.10}) contains six parameters, three from $g$ and the three
$f_\alpha$'s.  These, along with the three gauge parameters, [not
displayed in (\ref{eq:3.10})],  give the requisite number of nine
parameters.  The Abelianization of the Chern-Simons form via $a$ works only in
regions where $\sum f^2_\alpha \ge 2 f_1 f_2 f_3$, so that the square
root is well defined.  If this is not the case, one needs to use the
absolute value of $\sum f^2_\alpha - 2 f_1 f_2 f_3$ to get a real $a$.
This can lead to some nonanalyticity in $a$ as a function of the spatial
coordinates.

\section{Towards a Non-Abelian Fluid Mechanics}

We now turn to the question of whether our results can be used in a
physical context.  First of all, there has recently been renewed interest
in general parameterizations of gauge fields, with the hope that the low
energy physics of gauge theories might be clearer in certain cleverly
chosen parameterizations \cite{faddeev}.  Our work certainly fits in with this general
philosophy.

Secondly, notice that the Clebsch parameterization and the consequent
reduction of the Chern-Simons form are very useful in analyzing the evolution of
magnetic helicity \cite{JP1}.
Considerations of a non-Abelian analogue of magnetic helicity, which
may be relevant in the symmetry restored phase of the standard
electroweak theory, for example, can be significantly aided by our
analysis.

However, we now turn to a possible third application of our results: the construction of
non-Abelian fluid mechanics that may be relevant to the analysis of 
collective modes in the quark-gluon plasma.
The free Hamiltonian for nonrelativistic fluid mechanics is given by
\beq
H= \int \rmd^3r~\half~ \rho {\bf v}^2 
\label{eq:4.1}
\eeq
where $\rho$ is the matter density field and ${\bf v}$ is the velocity field.
The free evolution equations that these quantities satisfy are
\beqar
{\partial \rho \over \partial t} + {\bf \nabla} \cdot (\rho {\bf v})&&=0\nonumber\\
{\partial {\bf v} \over \partial t} +({\bf v}\cdot {\bf \nabla}) ~{\bf v}&&=0
\label{eq:4.2}
\eeqar
The first is the continuity equation linking the current ${\bf j}\equiv \rho {\bf v}$
to the density; the second is the free Euler equation, stating that the acceleration 
vanishes. These equations can be obtained by Poisson bracketing with $H$, provided
the nonvanishing brackets for $\rho , {\bf v}$ are
\begin{mathletters}%
\beqar
\left\{ v_i ({\bf r }), \rho ({\bf r} ') \right\} &&=
{\partial \over \partial r^i} \delta ({\bf r}-{\bf r}')\label{eq:4.3a}\\
\left\{ v_i ({\bf r} ), v_j ({\bf r}') \right\} &&= -{\omega_{ij} ({\bf r} )\over
\rho ({\bf r} )} \delta ({\bf r}-{\bf r}')\label{eq:4.3b}
\eeqar
\end{mathletters}%
where $\omega_{ij} = \partial_i v_j -\partial_j v_i$ is the 
vorticity~\cite{ref:4new}.
A natural question is whether there exists a canonical 1-form
and a corresponding symplectic 2-form that give the Poisson bracket
algebra (\ref{eq:4.3a},b).
The {\it raison d'\^{e}tre} for the Clebsch
parameterization (for the vector field ${\bf v}$)
is to provide this canonical formulation of 
fluid mechanics \cite{hydro}.  
One verifies that if ${\bf v}$ is presented as
\beq
{\bf v}= {\bf \nabla} \theta ~+~\alpha ~{\bf \nabla} \beta ,
\label{eq:4.4}
\eeq
where the canonical pairs of variables are identified as $(\rho ,\theta )$
and $(\rho \alpha , \beta )$, i.e.,
\begin{mathletters}%
\beqar
\left\{ \theta ({\bf r} ), \rho ({\bf r} ')\right\}&&= \delta ({\bf r}-{\bf r}')\label{eq:4.5a}\\
\left\{ \beta ({\bf r} ), \rho \alpha ({\bf r} ')\right\}&&=
\delta ({\bf r}-{\bf r}')\label{eq:4.5b}
\eeqar
\end{mathletters}%
then the algebra (\ref{eq:4.3a},b) is reproduced.\footnotemark[1]
\footnotetext[1]{\baselineskip=12pt The following observation allows
understanding the need of the Clebsch
parameterization for a canonical formulation. Although the algebra (\ref{eq:4.3a},b)
satisfies the Jacobi identity, it is nevertheless singular in that the Chern-Simons form
constructed from the velocity
\begin{displaymath}
CS ({\bf v}) = \int \rmd^3r ~ \epsilon^{ijk} v_i \partial_j v_k
\end{displaymath}
Poisson commutes with $\rho$ and ${\bf v}$; in other words, the Poisson algebra
(\ref{eq:4.3a},b) has a kernel given by the Chern-Simons form for the velocity.
This is an obstruction to the construction of a symplectic 2-form. The obstruction
is removed when ${\bf v}$ is taken in the Clebsch parameterization, for then the
Chern-Simons density is a total derivative and the Chern-Simons integral becomes
a surface term with no bulk contribution.}

A Lagrangian that incorporates the canonical 1-form and the parameterization
(\ref{eq:4.4}) is 
\beq
L= \int \rmd^3r~\left[ \half \rho {\bf v}^2 +\theta \left( {\dot \rho} +
{\bf \nabla} \cdot (\rho {\bf v}) \right) -\rho \alpha
\left( {\dot \beta} +{\bf v}\cdot {\bf \nabla} \beta \right)
\right]\label{eq:4.6}
\eeq
The time-derivative terms (denoted by the over-dot) supply the 1-form, 
while variation with respect to ${\bf v}$
yields (\ref{eq:4.4}).
Apart from total derivatives, $L$ can also be presented as
\beq
L= \int \rmd^3r~ \left[ -\rho \left( {\dot \theta} +\alpha {\dot \beta}\right)
-{\bf j} \cdot \left( {\bf \nabla} \theta + \alpha {\bf \nabla}\beta
\right) + {{\bf j}^2 \over 2 \rho}\right]
\label{eq:4.7}
\eeq
where we now use ${\bf j}$ instead of $\rho {\bf v}$ and the Clebsch variables have
been clearly exposed. Since the construction (\ref{eq:1.3}) of an Abelian vector from an $SU(2)$
pure gauge presents that vector in Clebsch form, we may replace (\ref{eq:4.7})
by
\begin{equation}
L=  \int \rmd^3r~\left[ 2 j^\mu  \tr (t^3 g^{-1} \partial_\mu g)- \sqrt{j^\mu j_\mu }~\right]
\label{eq:4.8}
\eeq
where now the kinetic term has also been generalized to the relativistic expression
and $j^\mu \equiv (\rho ,{\bf j})$ \cite{jpoly}. In the nonrelativistic limit
\beq
-\int \rmd^3r~ \sqrt{j^\mu j_\mu} \equiv -\int \rmd^3r~\sqrt{\rho^2 -{\bf j}^2} 
 \longrightarrow -\int \rmd^3r~ \rho +\int \rmd^3r~ {{\bf j}^2\over 2\rho}\label{eq:4.9}
\eeq
(The contribution $\int \rmd^3r~\rho$ to $L$ is immaterial; it is a constant of motion.)

The formula (\ref{eq:4.8}) suggests a non-Abelian generalization. $j^\mu$ is promoted
to an index-carrying ``color" current, $j^{\alpha\mu}$, and it is coupled to a
non-Abelian, ``Clebsch parameterized", vector constructed as in (\ref{eq:2.5}).
\beq
L= \int \rmd^3r~ \left[ -{1\over a} j^{\alpha\mu} \tr (I^\alpha g^{-1}\partial_\mu g)
- \sqrt{j^{\alpha \mu} j^\alpha_\mu}~\right]
\label{eq:4.10}
\eeq
Interaction with a dynamical gauge field can be included by promoting the derivative of
$g$ to a gauge-covariant derivative, gauged on the right, i.e.,
\beq
L= \int \rmd^3r~ \left[ -{1\over a} j^{\alpha\mu} \tr (I^\alpha g^{-1}D_\mu g)
- \sqrt{j^{\alpha \mu} j^\alpha_\mu}~\right] -\quar\int \rmd^3r~F^{\alpha \mu\nu}
F^\alpha_{\mu\nu}
\label{eq:4.11}
\eeq
with
\begin{equation}
D_\mu g = \partial_\mu g - e g A_\mu
\label{eq:4.12}
\end{equation}
$A_\mu = ~A^\alpha_\mu I^\alpha$ are independent non-Abelian gauge potentials (not given by $g$)
leading to the field strengths $F^\alpha_{\mu\nu}$.
The gauge transformation properties are
\begin{equation}
\begin{array}{cc}
\displaystyle g' = gh \ , &\qquad
\displaystyle A' = h^{-1}Ah + \frac{1}{e} h^{-1}
\rmd h \\[2ex]
\displaystyle j'_\mu = h^{-1} j_\mu h
\end{array}
\label{eq:4.13}
\end{equation}
where $j_\mu =j^\alpha_\mu I^\alpha$.

We expect that the Lagrangian (\ref{eq:4.11}) will describe non-Abelian
magnetohydrodynamics, namely the dynamics of a fluid with
non-Abelian charge coupled to non-Abelian fields.
The current density will be 
${\bf j}^\alpha$ as given by its
equation of motion. This 
gluon hydrodynamics can be useful for non-Abelian plasmas, such as the
quark-gluon plasma. Details of (\ref{eq:4.11})  and possible applications
are under further study. In a related investigation, conventional
fluid mechanics is generalized so that it enjoys a supersymmetry~\cite{jpoly2}.
\vskip .1in
\noindent{\bf Acknowledgments}

Conversations with S. Deser and A. Polychronakos initiated the research by one of us (RJ), who
also benefited from the suggestion by J. Mickelsson that a promising route to a
non-Abelian Clebsch parameterization could be through group-theoretic reduction
of the Chern-Simons term. V.P.N thanks I.~Singer for useful comments.

\end{document}